\documentclass[epj,twocolumn]{webofc}
\usepackage[varg]{txfonts}   
\woctitle{Hadron Collider Physics symposium 2012}
\usepackage{amsmath}	

\usepackage[usenames]{color}

\begin{document}
\title{
Production and decay of a heavy radion in Randall-Sundrum model at the LHC
}
%
%

\author{Yoshiko Ohno\inst{1}
\and
        Gi-Chol Cho\inst{2}
}

\institute{
Graduate School of Humanities and Sciences,
Ochanomizu University, Tokyo 112-8610, Japan
\and
Department of Physics, Ochanomizu University, Tokyo 112-8610, Japan
          }

\abstract{%
We study production and decay of a radion predicted in the
Randall-Sundrum model at the LHC. 
The radion is a scalar particle and its production and decay channels 
are quite similar to those of the Higgs boson. 
We find constraints on the model parameter space taking account of the
results of the Higgs boson search at the LHC. 
We also study a possibility discriminating the radion from the 
scalar particles in the other model such as the supersymmetric standard
model at the LHC experiments. 
}
\maketitle
\section{Introduction}
\label{intro}
A warped extra dimension model proposed by Randall and Sundrum (RS) 
\cite{Randall:1999ee} is one of the promissing candidates of physics beyond the Standard 
Model (SM) to solve the gauge hierarchy problem.  
The RS model is formulated in the five-dimensional anti de-Sitter 
(${\rm AdS}_5$) space-time, and 
the extra (fifth) dimension is compactified on an orbifold $S_1/Z_2$ of
a radius $r_c$. 
In the RS model, there are two 3-branes located at two orbifold fixed points
on the coordinate of the fifth dimension $y=0$ and $y=\pi r_c$. 
The former is called a hidden brane while the latter is called a  
visible brane. 
All the SM particles are confined in the visible brane 
and only the 
gravity is allowed to propagate into the fifth dimension. 
With this set-up, the hierarchy between the electroweak scale and the
Planck scale is understood reasonably if the distance between two 
3-branes (i.e., compactification radius $r_c$) and the AdS curvature 
($k$) 
satisfies a condition, $k r_c = 10 \sim 12$. 
It is, therefore, necessary to stabilize the radius of the extra
dimension to solve the gauge hierarchy problem. 
A simple stabilization mechanism of the distance between two branes  
is known as the Goldberger-Wise mechanism \cite{Goldberger:1999uk}. 
A radion is a scalar field which corresponds to the fluctuation of
the distance between two 3-branes. After the stabilization of the 
compactification scale, 
the radion has a finite mass and couples to the trace of the 
energy-momentum tensor of the SM. As a result, production and decay 
channels of radion at the hadron collider are the same with those of
Higgs boson in the SM. 
In this presentation, we show results of our study about production and 
decay of the radion at the LHC. 
We find constraints on the mass and coupling of the radion from the 
results of the Higgs boson search at the LHC.  
We also examine a possibility to discriminate the radion from heavier 
Higgs bosons in the minimal supersymmetric standard model (MSSM). 
We found that, if the mass spectrum of Higgs bosons in the MSSM follows 
so called the ``decoupling scenario'', the decay channels into the 
weak gauge boson pair ($WW$ and $ZZ$) are useful channels to 
distinguish the radion in the RS model from the heavier Higgs bosons 
in the MSSM at the LHC. 

\section{Production and decay of the radion}
\label{sec-2}
The radion couples to the trace of the energy-momentum tensor of 
the SM. 
The interactions of radion with SM fields are described by the following 
Lagrangian: 
\begin{eqnarray}
 \mathcal{L}_{\rm{int}}&=&\frac{\phi}{\Lambda_\phi}
\left\{
T^\mu_\mu({\rm SM}) + T^\mu_\mu(\rm{SM})^{\rm{anom}}
\right\}, 
\label{lag}
\end{eqnarray}
where $\phi$ is the radion field and $\Lambda_\phi$ is its vacuum
expectation value (VEV). 
The trace of energy-momentum tensor $T^\mu_\mu(\rm{SM})$ is 
given by 
\begin{eqnarray}
T^\mu_\mu({\rm SM})&=&\sum_{f} m_f \bar{f}f-2m_W^2 W^+_\mu 
W^{-\mu}-m_Z^2 Z_\mu Z^\mu
\nonumber \\
&&~~~
+(2m_h^2h^2-\partial_\mu h \partial^\mu
h)+ \cdots 
\label{tmn1}
\end{eqnarray}
while $T^\mu_\mu({\rm SM})^{\rm anom}$ comes from the scale anomaly  
\begin{eqnarray}
T^\mu_\mu({\rm SM})^{{\rm anom}}=\sum_a
\frac{\beta_a(g_a)}{2g_a}
F^a_{\mu\nu}F^{a\mu\nu},  
\label{tmn2}
\end{eqnarray}
where $a$,  $\beta_a$ and $F_{\mu\nu}^a$ are 
the gauge index, the $\beta$-function and the field strength 
for a corresponding gauge group, respectively~
(for detail, see \cite{Csaki:2007ns,Cheung:2000rw}).

\begin{figure}[htbp]
\centering
\vspace{0cm}
\includegraphics[width=7cm,clip]{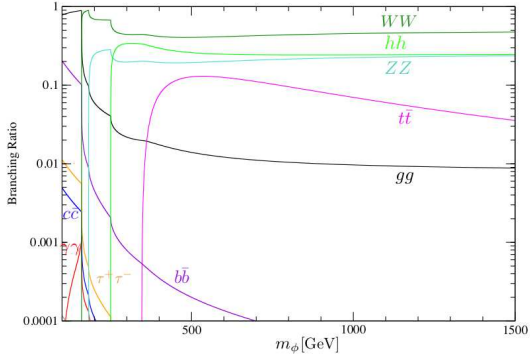}
\vspace{0cm}       
\caption{Branching ratio of the radion decaying to the SM particles 
as a function of the radion mass $m_\phi$. 
For a decay to the Higgs boson pair, $\phi \to h h$, the Higgs boson 
mass is fixed at $125~{\rm GeV}$. }
\label{radionBR}       
\end{figure}

From eqs.~(\ref{lag}), (\ref{tmn1}) and (\ref{tmn2}), we can calculate
the decay rate of the radion into the SM fermions, gauge bosons and the
Higgs boson. 
In Figure~\ref{radionBR}, we show the branching ratios of the radion 
decaying to the various SM particles as a function of the radion mass 
$m_\phi$. 
The Higgs boson mass $m_h$ in the final state is fixed at $125~{\rm
GeV}$. 
Note that the dominant decay modes of radion are the $WW$, $ZZ$ and 
$hh$ channels for $m_\phi \ge 500~{\rm GeV}$. 

\section{Constraints on the radion mass}
\label{sec-3}
The RS model is characterized by two parameters, $m_\phi$ and 
$\Lambda_\phi$.  
Constraints on the parameter $\Lambda_\phi$ has been studied indirectly
using the result of the 1st Kaluza-Klein graviton search at the 
LHC~\cite{Tang:2012uw}.  
As mentioned in a previous section, the decay modes of the radion are 
completely same with those of the SM Higgs boson up to the couplings.  
In this section, therefore, we study constraints on $m_\phi$ and 
$\Lambda_\phi$ using the experimental data of the Higgs boson search at 
the LHC. 

At the LHC, a signal of Higgs-like particle has been found at about
$125~{\rm GeV}$ in both ATLAS and CMS, and the wide mass region except
for $\sim 125~{\rm GeV}$ has been excluded. 
For example, the ATLAS experiment excludes the mass smaller than
$558~{\rm GeV}$ from the decay channels $h\to WW$ and $h\to
ZZ$~\cite{:2012va}.  
This mass bound may be understood as a lower mass bound on an 
scalar particle in some new physics models if the scalar particle 
couples to $WW$ and $ZZ$. 
Since the radion also decays into both $WW$ and $ZZ$, we examine 
constraints on the mass of the radion $m_\phi$ and the coupling 
$\Lambda_\phi$ from the results of ATLAS experiment. 
In the following, we study the allowed parameter space of the radion 
under the condition 
\begin{eqnarray}
&&
 \sigma(pp\to h)\times \text{Br}(h \rightarrow WW/ZZ)\Big|_{m_h=558~{\rm
 GeV}}
~~~~~~~~~~
\nonumber \\ 
&&~
 \geq
\sigma(pp\to \phi)\times \text{Br}(\phi\rightarrow
 WW/ZZ). 
\label{CSBR}
\end{eqnarray}

\begin{figure}[htbp]
\centering
\includegraphics[width=6cm,clip]{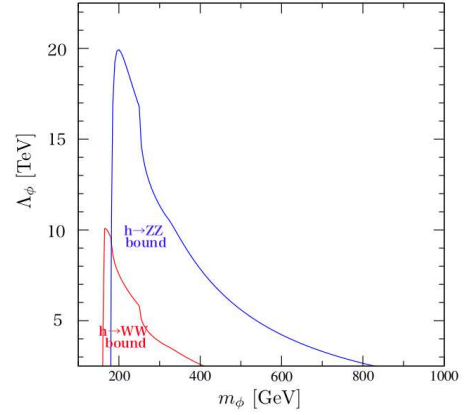}
\vspace{0cm}       
\caption{Constraints on $m_\phi$ and $\Lambda_\phi$ from 
the results of the Higgs search experiment at the LHC. 
The red and blue contours are obtained from the experimental bounds on 
the decay channels $h\to WW$ and $h\rightarrow ZZ$, respectively. 
}
\label{contour}       
\end{figure}

We show constraints on $m_\phi$ and $\Lambda_\phi$ taking account of 
the condition (\ref{CSBR}) in Figure~\ref{contour}.  
Two contours correspond to the 95\% CL
 bounds obtained from the Higgs search limit in the $WW$  
channel (red) and the $ZZ$ channel (blue), respectively. 
The inner of contours are excluded region. 
It can be seen from the figure that the radion mass $m_\phi$ should be
larger than $500~{\rm GeV}$ for $\Lambda_\phi < 5~{\rm TeV}$. 
As $\Lambda_\phi$ increases, bound on $m_\phi$ is weakened. 
Note that constraints from $h\to ZZ$ are much stronger than those 
from $h\to WW$. 
For example, $m_\phi=300~{\rm GeV}$ and $\Lambda_\phi=8~{\rm TeV}$ is
strongly disfavored by the experimental bounds for $h\rightarrow ZZ$ 
(not by $h\to WW$) at the LHC. 
However, the sensitivity of the $WW$ mode for small $m_\phi$ is slightly 
better than $ZZ$ mode. 

From this result, we conclude that the radion lighter than $\sim
300~{\rm GeV}$ is disfavored from the Higgs search experiment at the LHC 
unless the parameter $\Lambda_\phi$ is larger than $10~{\rm TeV}$. 
Since too large $\Lambda_\phi$ is unfavorable because it may lead to 
a new hierarchy between the weak scale and $\Lambda_\phi$, 
we focus on the heavy radion search at the LHC. 
%

\section{Radion vs. MSSM Higgs}
\label{sec-4}
Once a new scalar particle beyond the SM is found in the future LHC
experiments, we must study if it is the radion or not, because  
various models predict an existence of neutral scalar particles. 
In this section, we examine a possibility to discriminate the radion 
from the heavier Higgs bosons in the MSSM as an example. 
In the MSSM, there are five Higgs bosons -- two $CP$-even scalars ($h,H$),  
one $CP$-odd scalar $(A)$ and two charged scalars ($H^\pm$). 
It is known that the mass of scalar particles discovered at the LHC,
$\sim 125~{\rm GeV}$, is very close to the upper limit of the lightest 
Higgs boson $h$ in the MSSM. 
A possible explanation on the mass of $h$ is known as so called 
``decoupling 
scenario'' where the lightest Higgs boson $h$ has almost same property
with the SM while the other four Higgs bosons are much heavier than $h$
and degenerate in mass. 
In the following, we identify the lightest Higgs boson $h$ as a particle 
discovered at the LHC and compare the production and decay of the heavy
neutral Higgs bosons ($H, A$) with those of the radion. 
\begin{figure}[htbp]
\centering
\includegraphics[width=7cm,clip]{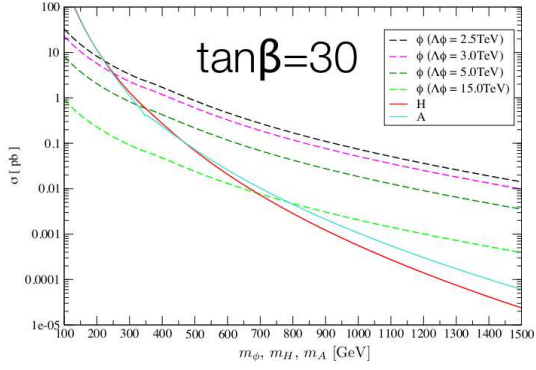}
\vspace{0cm}       
\caption{Production cross sections 
of $\phi,H,A$ as functions of their mass. 
}
\label{cs30}       
\end{figure}

Figure~\ref{cs30} shows the production cross sections of the radion 
$\phi$ and MSSM Higgs bosons $H,A$ at the LHC as functions of their  
mass. 
The dashed lines represent the radion with various values of the 
parameter $\Lambda_\phi$ from $2.5~{\rm TeV}$ to $15~{\rm TeV}$. 
The solid lines in red and in cyan correspond to $H$ and $A$,
respectively.  
The cross sections of $H$ and $A$ are obtained for $\tan\beta=30$ as an
example, where $\tan\beta$ is a ratio of VEV of two Higgs doublets. 
It is easy to see that,  
when the mass is larger than $500~{\rm GeV}$, the production cross
section of the radion is much larger than those of $H$ and $A$ for
$\Lambda_\phi \le 5~{\rm TeV}$.  

A distinctive feature of the decoupling scenario can be found in the
interactions of the Higgs bosons.  
The interactions of the heavier Higgs bosons $H,A$ to the weak gauge
bosons $W,Z$ are proportional to $\cos(\alpha-\beta)$, where $\alpha$ is
the mixing angle of the mass matrix of the $CP$-even Higgs bosons. 
The decoupling scenario is realized when $\beta \simeq \alpha + \pi/2$ 
so that the interactions of $H$ and $A$ to $WW$ and $ZZ$ are highly
suppressed. 
On the other hand, the interactions of radion to $WW$ and $ZZ$ have no 
suppression like $H$ and $A$. 
We, therefore, expect relatively larger decay rates of the radion to
$WW$ and $ZZ$ as compared to those of $H$ and $A$, and these decay 
modes could be promissing channels to distinguish the radion from the 
MSSM Higgs bosons.  
We calculated the number of events $N$ of the processes $pp \to \phi \to 
WW/ZZ$ at the LHC with the center-of-mass energy $\sqrt{s}=7~{\rm TeV}$
and the integrated luminosity $\int dt {\cal L}=15~{\rm fb}^{-1}$. 
The results are shown in Table~\ref{tab-1} for $\Lambda_\phi=8~{\rm
TeV}$ and $15~{\rm TeV}$. 
In the table, the number of events of the $\phi\to ZZ$ mode 
is absent for $m_\phi=200~{\rm GeV}$. 
This is because that the radion mass $m_\phi=200~{\rm GeV}$ is 
disfavored from the search of Higgs boson in the $h\to ZZ$ channel 
at the LHC as shown in Figure \ref{contour}. 
We obtain the number of events more than a thousand (a few hundred) 
for $m_\phi=200~{\rm GeV}$ and $\Lambda_\phi=8~{\rm TeV}~(15~{\rm
TeV})$. 
Even for larger mass, $m_\phi=500~{\rm GeV}$, we obtain $O(10)$ events 
for both $\Lambda_\phi=8~{\rm TeV}$ and $15~{\rm TeV}$ in both the $WW$
and $ZZ$ channels. 
Note that, as already discussed, the decays into $WW$ and $ZZ$ of
heavier Higgs bosons ($H$ and $A$) in the decoupling scenario of MSSM 
are strongly suppressed.   
Thus, when a heavy and neutral scalar particle is discovered in 
the future LHC experiment, we can expect to test a possibility of the
radion using the $WW$ and $ZZ$ channels. 
\begin{table}[htbp]
\begin{tabular}{llll}
\hline
$m_\phi$ & 200\,GeV& 500\,GeV& 1\,TeV \\ 
 $\Lambda_\phi=8$\,TeV&&& \\ \hline 
$\sigma\times \rm{Br}$ $(\phi\rightarrow WW)$ & $1.9\times10^3$ &
	 $6.6\times 10$& $3.2$ \\
$\sigma\times \rm{Br}$ $(\phi\rightarrow ZZ)$ & --------- &
	 $3.2\times 10$& $1.6$ \\
\hline \hline 
$\Lambda_\phi=15$\,TeV&&& \\  \hline
$\sigma\times \rm{Br}$ $(\phi\rightarrow WW)$ & $5.4\times10^2$ &
	 $1.9\times 10$& $9.0\times10^{-1}$ \\ \hline
$\sigma\times \rm{Br}$ $(\phi\rightarrow ZZ)$ & --------- &
	 $9.0$& $4.6\times10^{-1}$ \\
\hline
\end{tabular} 
\caption{Number of events of the radion decay $\phi\rightarrow WW/ZZ$ 
at $\sqrt{s}=7$\,TeV for the integrated  luminosity 15\,$\rm{fb}^{-1}$.}
\label{tab-1}       
\end{table}
\section{Summary}
\label{summary}
We have shown the results of our recent study for the 
production and decay of the radion in the RS model at the LHC. 
We first found constraints on the radion mass $m_\phi$ and the 
coupling $\Lambda_\phi$ from the experimental results of search for the 
Higgs boson at the LHC, since the radion has same decay channels with
the Higgs boson.  
We examined a possibility to distinguish the radion from the other
scalar particles such as the heavier Higgs bosons in the MSSM. 
The discovery of the Higgs-like particle at the LHC tells us that 
the decoupling scenario on the Higgs boson masses in the MSSM is one of 
the possible scenarios of the MSSM. 
We discussed that the decays into $WW$ and $ZZ$ are promissing 
channels to discriminate the radion from the heavier Higgs 
bosons $H,A$ in the decoupling scenario of the MSSM. 
When the radion mass is $500~{\rm GeV}$, a few $\times 10$ events are
expected in the radion decays to $WW$ and $ZZ$ while decays of $H$ and
$A$ into those final states are highly suppressed in the decoupling
scenario of the MSSM. 

\vspace{0.5cm}

The work of G.C.C is supported in part by Grants-in-Aid for Scientific 
Research from the Ministry of
Education, Culture, Sports, Science and Technology (No.24104502) and 
from the Japan Society for the Promotion of Science (No.21244036).

\end{document}